\documentclass[12pt,a4paper]{article}
\usepackage{amsfonts}
\usepackage{amsmath}
\usepackage{epsfig}
\usepackage{graphicx}

\begin{document}
\title{ \bf  Initial problem for heat equation with multisoliton inhomogeneity and one-loop quantum corrections}

\author{ S. Leble, A. Yurov,\\
\small
Faculty of Applied Physics and Mathematics\\
\small Technical University of Gda\'{n}sk,  \\
\small  ul. G.Narutowicza, 11/12 80-952, Gda\'{n}sk-Wrzeszcz, Poland,\\
\small   email leble@mif.pg.gda.pl\\
\small and\\
\small Kaliningrad State University, Theoretical Physics Department,\\
\small Al.Nevsky st.,14, 236041 Kaliningrad, Russia.\\}

\maketitle
 \begin{abstract}
 The generalized
zeta-function is built by a dressing method based on the Darboux
covariance  of the heat equation and used to evaluate the
correspondent functional integral in quasiclassical approximation.
Quantum corrections to a kink-like solutions of Landau-Ginzburg
model are calculated.
\end{abstract}
\section{Introduction}
In the paper of V.Konoplich \cite{Ko} quantum corrections to a few
classical solutions by means of Riemann zeta-function are
calculated. Most interesting of them are the corrections to the kink
- the separatrix solution of field $\phi^4$ model. The method of
\cite{Ko} is rather complicated and it could be useful to simplify
it. We use the dressing technique based on classical Darboux
transformations (DT) with a new applications to Green function
construction \cite{LeZa3}.  It is the main aim of this note with
eventual possibility to generalize the result due to universality of
the technique when a link to integrable (soliton, SUSY) systems is
established \cite{Suk04}. The suggested approach open new
possibilities; for example it allows to show the way to calculate
the quantum corrections to Q-balls \cite{CE} and periodic solutions
of the models. The last problem is posed in the useful review
\cite{TD}.
\medskip

\section{Heat equation Cauchy problem}
We will base on the DT-covariance of the heat equation for the
function $\rho(\tau, x, y)$
\begin{equation}\label{heat}
    -\rho_{\tau}+\rho_{xx} + u(x)\rho = 0,
\end{equation}
that means the form-invariance of (\ref{heat}) with respect to
iterated DT, defined by the Wronskian $W[\phi_1,...,\phi_N]$ of the
solutions of (\ref{heat})
\begin{equation}\label{Wro}
    \begin{array}{c}
      \rho \rightarrow \rho[N]=\frac{W[\phi_1,...,\phi_N,\rho]}{W[\phi_1,...,\phi_N]}, \\
      u \rightarrow u[N] = u + 2ln_{xx}W[\phi_1,...,\phi_N]. \\
    \end{array}
\end{equation}

Consider now a Cauchy problem for the equation (\ref{heat}), where
$u(x)$ represents the reflectionless potential \cite{{NMPZ}} in a sense that it could be produced by the DT and
the initial condition is
\begin{equation}\label{delt}
    \rho(0,x,y) = \delta(x-y).
\end{equation}
The problem is formulated for a Green function: it is rather general
and may be applied as a  model of  classical diffusion or heat
conductivity. We, however, would follow other applications in the
theory of quasiclassical quantization, where the function $\rho$ is
treated as density matrix whence $\tau$ stands for inverse
temperature \cite{DHN}.

The algorithm of such problem solution is the dressing procedure
organized by a sequence of DTs defined by (\ref{Wro}):
\begin{equation}\label{seq}
\begin{array}{c}
  (\frac{\partial}{\partial x} - \ln_x\phi_1(x,y))\rho_0(0,x,y) = g_1(x,y), \\
 (\frac{\partial}{\partial x} - \ln_x\phi_2[1](x,y))g_1(x,y) = g_2(x,y),..., \\
 (\frac{\partial}{\partial x} - \ln_x\phi_k[k-1](x,y)) g_{k-1} = g_k{x,y},\\
 g_N(x,y) = \delta(x,y), 2 \leq k \leq N.
\end{array}
\end{equation}
and the following theorem

{\bf Theorem} The function $\rho[N]$ being built by (\ref{Wro}) will
be a solution of the problem (\ref{heat},\ref{delt}) with the
potential $u[N]$, if $\rho(\tau,x,y)$ is a solution of the
(\ref{heat}) with the initial condition $\rho_0(0,x,y)$.

The result is  used when static solutions of $\phi^4$ model are
quantized  by means of Riemann function $\zeta(s)$  \cite{Ko}
expressed via the Green functions of the (\ref{heat}) (see also
\cite{LeZa3}). The one-loop quantum correction to action is
evaluated directly as
$$
S_q = - \zeta'(0).
$$

 \section{ Example of kink }

 Most popular example of the kink is obtained in this
scheme by means of DT over zero seed $u=0$. The solution $\rho$ of
(\ref{heat}) with $\rho_0$ as initial condition for this case is a
simple heat equation solution
\begin{equation}\label{hesol}
   \rho(\tau,x.y) =
   \frac{1}{2\sqrt{\pi\tau}}\int_{-\infty}^{\infty}\rho_0(z,y)\exp[-(x-z)^2/4\tau]dz.
\end{equation}
The initial condition $\rho_0$ is evaluated by direct integration in
(\ref{seq}):
\begin{equation}\label{po0}
    \rho_0(x,y) = \phi_1(x)  \left\{\begin{array}{ll}
\phi_1^{-1}(y)\;,& x>y \\
      0, &x<y \\
    \end{array}
    \right.
\end{equation}
The Green function $\rho[2]$ (density matrix) for the kink solution
as the potential is built by the two-fold DT by the Wronskian
formula (\ref{Wro}) that results in
\begin{equation}\label{Gtt}
\begin{array}{c}
 \rho[2](\tau,x,y) =
\exp[\frac{-(x-y)^2}{4 \tau}]/2\sqrt{\pi \tau} + \\
  +\frac{1}{2}\sum_{m=1}^{2}\rho_m\psi_m(x)\psi_m(y)[Erf [\frac{(x-y+2 b_m
  \tau)}{2\sqrt{\tau}}]
  -Erf[\frac{(x-y-2 b_m \tau)}{2\sqrt{\tau}}],
\end{array}
\end{equation}
where $b_k = km/\sqrt{2}
$, $\rho_k=||\psi||^{-2}$, k=1,2. After multiplication of the Green function by $\exp[-4m^2\tau]$:
$$
\rho \rightarrow \rho \exp[-4m^2\tau],
$$
the first term of the Green function leads to a divergent integral.
This divergency is well-known, its origin is a zero vacuum
oscillations.In our approach this fact has transparent explanation,
because the divergent term is simply a solution of heat equation
with constant coefficients, that appear when the self-action of
scalar field is neglected. Such divergence is usually compensated by
addition of contra terms of normal order.

Our procedure deletes all
ultraviolet divergencies of 1+1 $\phi^4$ model
 including energy of zero oscillations and one-meson states if one
evaluates the generalized zeta-function by the formula
\begin{equation}\label{zeta}
    \zeta_D(s) = M^{2s}\frac{\int_0^\infty \gamma(t)t^{s-1}dt}{\Gamma(s)}
\end{equation}
$\Gamma(s)$ is the Euler gamma function and M is a mass scale. The
function $\gamma(t)$ in  the   integrand of (\ref{zeta}) is
expressed via the Green functions
 $G(x,y,\tau)$ and
   $G_0(x,y,\tau)$ difference. The result coincides with one from \cite{Ko}.

\section{Conclustion}

   As a conclusion let us note that
this approach is elaborated in \cite{LY} (published in a local
conference abstract book) and allows to calculate one-loop
corrections to the N-level reflectionless potential and, very
similarily,  solitons of SG. Some eventual applications are visible
in the case studied at \cite{TMCD}.


\begin{thebibliography}{bog-kon98-1}
\bibitem{Ko} Konoplich R.V. Quantum corrections calculations to nontrivial
classical solutions via zeta-function. TMP,1987,v73,p 379-392. The
zeta-function method in field theory at finite temperature.(Russian)
 Teoret. Mat. Fiz. 78 (1989), no. 3, 444--457; translation in Theoret. and Math. Phys. 78 (1989),
  no. 3, 315--325.  One-loop quantum corrections to the energy of extended objects. Nuclear Phys.
   B 323 (1989), no. 3, 660--672.

\bibitem{LeZa3} Leble S., Zaitsev A, The Modified Resolvent for the One-dimensional Schrodinger Operator
with a reflectionless potential and Green Functions in
Multidimensions (with  ){\it J.Phys. A:Math.Gen.} v.28 (1995)
p.L585-L588.

\bibitem{Suk04} Sukumar, C. V. Green's functions, sum rules and matrix
elements for SUSY partners. J. Phys. A 37 (2004), no. 43,
10287--10295.
\bibitem{CE} Cervero J.M, Estevez P.G. Exact two-dimensional Q-balls near
the kink phase Phys.Lett.B,v176,p139-142,1986.

\bibitem{TD} Tuszy\'nski, J. A.; Dixon, J. M.; Grundland, A. M.
Nonlinear field theories and non-Gaussian fluctuations for near-critical many-body systems.
 Fortschr. Phys. 42 (1994), no. 4, 301--337.

\bibitem{LY} Leble, S.B. and Yurov, A.V. (1993) On the quantum corrections to classical
solutions via generalized zeta-function, in \textit{Abstracts of
XXVII sci. conf. Kaliningrad State University, Kaliningrad}, p. 157.

\bibitem{NMPZ} Novikov S.P. Manakov S.V. Pitaevski L.P. Zakharov V.E.
Theory of Solitons. Plenum, New York, 1984.

\bibitem{TMCD} Tuszy\'nski, J. A.; Middleton, J.; Christiansen, P. L.; Dixon, J. M. Exact
eigenfunctions of the linear ramp potential in the Gross-Pitaevskii equation for the Bose-Einstein
condensate. Phys. Lett. A 291 (2001), no. 4-5, 220--225.
\end{thebibliography}
\end{document}